\documentstyle[preprint,eqsecnum,aps]{revtex}

\begin{document}
\draft
\title{Metastable Dynamics of the Hard-Sphere System}
\author{Joonhyun Yeo\cite{padd}}
\address{
The James Franck Institute
and Department of Physics \\
The University of Chicago,
Chicago, Illinois  60637}
\date{\today}
\maketitle
\begin{abstract}
The reformulation of the mode-coupling theory (MCT) of the
liquid-glass transition which incorporates
the element of metastability is applied to the hard-sphere
system. It is shown that the glass transition
in this system is not a sharp one at
the special value of the density or the packing fraction, which is in
contrast to the prediction by the conventional MCT. Instead
we find that the slowing down of the dynamics occurs over a range
of values of the packing fraction. Consequently, the exponents governing the
sequence of time relaxations in the intermediate time regime are
given as functions of packing fraction with one additional parameter
which describes the overall scale of the metastable potential energy for
defects in the hard-sphere system. Implications of the present model
on the recent experiments on colloidal systems are also discussed.
\end{abstract}

\narrowtext
\section{introduction}
\label{sec:1}

In Ref.~\cite{my2,ym}, the notion of metastability was used in the
reformulation of the mode coupling theory (MCT) \cite{leut,mct:gen}
of the liquid-glass
transition. It was shown that the temperature dependence of the transition
is smooth, as observed in recent experiments \cite{dixon,exp:rec},
showing no evidence for the special transition temperature
conventionally assumed in MCT. In this paper, we apply the basic picture
obtained in Ref.~\cite{my2,ym} to a more
realistic situation by considering the effects
of the structure of supercooled liquids on the slowing down of the
dynamics. For a simple system like hard-sphere fluids
considered here, we find that the observed slowing down of the dynamics
occurs over a range of values of the
packing fraction. This is in contrast to the prediction by the
conventional MCT \cite{conv:mct}
that there exists a critical packing fraction for the
transition. Consequently, we find that the exponent parameters
describing the relaxation
sequence in the intermediate time regime are given as smooth
functions of packing fraction with one additional parameter, which is related
to the existence of metastable defects in the system.

Much of the recent theoretical progress on the liquid-glass transition
problem has been inspired by the MCT \cite{leut,mct:gen}. It was
particularly successful in explaining the very elaborate sequence of
time relaxations which has been observed in
many experiments \cite{exp:seq1,exp:seq2}. The conventional interpretation
of this theory assumes
a sharp transition as the system approaches the critical temperature
or the critical density. Although the sharp nature of the transition
is smeared by the cutoff effect \cite{dm}, the notion of a critical
temperature or density at which the dynamics is arrested
still remains in the theory \cite{new:mct}.
For example, the exponents governing
the time relaxations are defined only at that particular
temperature or density.
This is, however, incompatible with
the recent experimental results \cite{dixon} showing a
more universal behavior
and smooth temperature dependences. Other recent experiments \cite{exp:rec}
also indicate that the existence of the well-defined
transition temperature is very doubtful.

The basic ingredient in the reformulation of MCT in Ref.~\cite{my2,ym} is the
introduction of the metastable defect variable and its coupling to the
density fluctuations in the glass transition problem. The
defect variable is supposed to be governed by a very long
time scale compared to the mass density and
the metastability is incorporated through a double-well
type potential energy for the defect variable. By investigating
the conditions that the defect variable must satisfy
in order to be on the critical
surface associated with MCT \cite{goe2},
the authors of Ref.~\cite{my2} were able to
obtain the observed slowing down without
having to adjust the temperature. The condition for
slowing down is given
instead by a self-consistent
condition between the low activation barrier for defects and its weak
coupling to the density fluctuation \cite{my2,ym}.
The temperature dependence of the
parameters describing the metastable potential for defects turns out
to be smooth. The exponents for the time relaxation sequences can be determined
for a range of temperatures showing
a smooth temperature dependence compatible with recent experiments.

In the analysis of those models, however,
the wavenumber dependences were suppressed as a simplifying
assumption. The static structure
factor for the density fluctuation $S(q)$ was treated in a very simple manner
by assuming that it is constant for $q<\Lambda$ and zero otherwise, where
$\Lambda$ is the large wavenumber cutoff.
In this paper, we generalize the previous model by considering
a model where the structure factor has a
more realistic wavenumber dependence.
We focus here on the dynamics
of hard-sphere fluids for which
an accurate approximate analytical treatment of $S(q)$ is available.
We restrict our analysis here to
the intermediate time regime before the primary relaxation
where much of the analysis can be carried out analytically.
A detailed MCT analysis has been carried out in Ref.~\cite{goe1}
for this time regime, which is very useful in our analysis here.
We note that in the previous model \cite{my2,ym} the temperature
was represented by the
parameter $\xi$ which is proportional to
$\Lambda^3$ as well as to the temperature.
The appearance of the large wavenumber cutoff in the temperature
parameter $\xi$ is directly related to the suppression of the wavenumber
dependence in the model. When one considers the wavenumber dependence
of $S(q)$,
however, we find that the role played by $\xi$ in the previous case
is replaced, in the case of hard-sphere
fluids, by the more physical parameter,
the packing fraction.

The model in Ref.~\cite{my2,ym} is based on the fluctuating nonlinear
hydrodynamics (FNH) for a set of slow variables
consisting of the mass density $\rho ({\bf x})$, the momentum density
${\bf g}({\bf x})$ and the defect density $n_d ({\bf x})$.
As discussed in Ref.~\cite{das},
the wavenumber dependence can be incorporated into the FNH formulation
through the terms in the effective Hamiltonian $F$ that contains
spatial derivatives of the variables. More generally,
one can include terms where the coefficients depend explicitly
on the position ${\bf x}$.
We construct $F$ such that the terms that
depend on the density fluctuation $\delta\rho$
yield the correct static structure factor $S(q)$ for the hard-sphere
system, where $\delta\rho =\rho -\rho_0$ with
the average mass density $\rho_0$. In the presence of the defect variable, we
must also consider the effect of the defect structure on the
dynamical slowing down. Unfortunately, a microscopic description
for the structure factors in the presence of defects is not
available even for a simple system like a hard-sphere fluid.
We assume here that
the local behavior of defects is similar to that of particles.
Thus we assume that
any possible effect of the defect structure is reflected in the coupling
term in $F$ between the density fluctuation and the defect fluctuation
$\delta n_d$, where $\delta n_d \approx n_d -\bar{n_d} $ \cite{fn:dnd}
with the metastable defect density $\bar{n_d}$.
This assumption is reasonable
since a vacancy is created whenever a particle moves to a
different position to make an interstitial.
In this paper we assume, as a first approximation, that the coupling
term in $F$ between $\delta\rho$
and $\delta n_d$ has the same spatial form as
the term quadratic in $\delta\rho$.
As we shall see below, this corresponds to choosing
the static defect-density correlation
function to be proportional to $s(q)-1$, where
$s(q)=S(q)/(m\rho_0 )$ is the
dimensionless form of the structure factor
$S(q)$
with $m$ the mass of the particle.
This quantity is just the static structure factor for the
density fluctuation excluding the self-correlation effect.
The potential energy for $n_d$ is chosen to have the same double well
potential as in Ref.~\cite{my2,ym}. The basic picture of Ref.~\cite{my2,ym}
is then retained in the sense that the slowing down of the dynamics
corresponds to the limit of the low activation barrier for defects and
the weak coupling.

As mentioned above, MCT successfully explains the observed
sequence of time relaxations. This sequence can be summarized
as follows:
After the microscopic time scale, the system goes into the
intermediate time regime where the density
auto-correlation function $\phi (t)$ exhibits
the power-law
decay, $\phi\sim f+A_1 t^{-a}$
followed by the von-Schweidler relaxation, $\phi\sim
f-A_2 t^b$. The system then crosses over to the primary
($\alpha$-) relaxation regime which can be described by
a stretched exponential decay $\phi\sim\exp\{ (
t/\tau_\alpha )^\beta\}$.

In the intermediate time regime,
a detailed analysis is available \cite{goe1}
for the general MCT equations including the wavenumber dependence.
It is shown there that all
microscopic structural details are summarized into
one exponent parameter $\lambda$, which
gives the power-law and the von-Schweidler exponents $a$ and $b$.
Thus, given a microscopic model,
the corresponding $\lambda$ can be calculated. In the present case,
we find, assuming the system is on the critical surface, that
$\lambda$ depends upon the parameter describing the overall
scale of the metastable
potential for defects as well as on the packing fraction.
This result that the exponent parameter $\lambda$
is given as a function of the packing fraction is conceptually
very different from the results of the earlier MCT calculations on the
hard-sphere system \cite{conv:mct,mct:hs,mct:hs2,mct:hs3},
where only the simplest contribution
from the density fluctuation to the dynamic viscosity is taken into account.
In those cases, the transition is controlled by the packing fraction
and thus $\lambda$ is defined only at the critical packing fraction
just as in a second order phase transition.
According to the picture presented here,
the transition can take place over a
range of packing fractions with different values of the exponents.

In Sec.~\ref{sec2}, we construct the FNH model including the
wavenumber dependence and present
the relevant results to our discussion of the slowing down of the
dynamics. Since the calculation follows closely
those in Ref.~\cite{my2,ym} for
the wavenumber independent model, we emphasize here only the important
results. In Sec.~\ref{sec3}, we present the results of numerical
analysis of the model. In Sec.~\ref{sec:conc}, we conclude with
a brief discussion.

\section{Model}
\label{sec2}

The FNH approach developed in Ref.~\cite{my2,ym} is based on
the generalized Langevin equations for the slow variables,
$\{\rho, {\bf g}, n_d\}$. As mentioned in the previous section, the
wavenumber dependence can be implemented through
general spatially-varying coefficients of terms in the effective Hamiltonian
$F[\rho, {\bf g}, n_d]$, which governs
the static property. In the present model, $F$ is given by
\begin{equation}
F[\rho, {\bf g}, n_d]=F_K [\delta\rho , {\bf g}]+F_u [\delta\rho ]
+F_c [\delta\rho ,n_d]+F_v [n_d] .
\label{f}
\end{equation}
Here $F_K$ is the usual kinetic energy term
$F_K [\rho, {\bf g}]=\int d^3 {\bf x}\;{\bf g}^2/(2\rho )$.
For the terms that are dependent on the density fluctuation only, we take
the Ramakrishnan-Yussouff form \cite{ry}
\begin{equation}
F_u [\delta\rho ]=\frac{k_B T}{m}\int d^3 {\bf x}\;
[\rho\log\frac{\rho}{\rho_0}-\delta\rho ]-\frac{k_B T}{
2m^2}\int d^3 {\bf x}d^3 {\bf x}^\prime \; \delta\rho
({\bf x})C({\bf x}-{\bf x}^\prime )\delta\rho ({\bf x}^\prime ),
\label{ry:eq}
\end{equation}
where $k_B T$ is the temperature, $m$ is the mass
and $C(x)$ is the Ornstein-Zernike direct
correlation function for the density
fluctuation \cite{oz}.
For the coupling term, we have in general
\begin{equation}
F_c [\delta\rho , n_d]=\int d^3 {\bf x}d^3 {\bf x}^\prime \;
n_d({\bf x})B({\bf x}-{\bf x}^\prime )\delta\rho ({\bf x}^\prime )
\label{fc}
\end{equation}
for some function $B({\bf x})$ to be specified later. We write the
Fourier transform of $B({\bf x})$ as $B({\bf q})\equiv\int d^3{\bf x}
\exp (i{\bf q}\cdot{\bf x})B({\bf x})\equiv B\cdot
b(q)$, where the constant $B$ represents a scale of the coupling energy
with the appropriate dimension for $B(q)$ and the dimensionless $b(q)$
contains the wavenumber dependence of $B(q)$.
In Eq.~(\ref{f}),
the potential energy part $F_v [n_d ]$ for $n_d$ is given by
the same double well potential $h(n_d )$ with the metastable
defect density $\bar{n}_d$ as in Ref.~\cite{my2,ym}:
$F_v [n_d]=\int d^3 {\bf x}\; h(n_d)$, where
$h^\prime (\bar{n}_d )=0$. We define for later use $\mu\equiv
\bar{n}_d^2 h^{\prime
\prime}(\bar{n}_d)$ and $\nu\equiv\bar{n}_d^3 h^{\prime\prime\prime}
(\bar{n}_d)$ which are parameters describing the shape of $h(n_d)$.
We note that, if $\mu\rightarrow 0$, the potential $h(n_d )$
develops an inflection point at
$n=\bar{n}_d$, which corresponds to the potential
with very shallow metastable
wells and low activation barriers for the defects.

The static correlation functions can readily be found from Eq.~(\ref{f}):
\begin{mathletters}
\begin{eqnarray}
&&\langle\delta\rho ({\bf q})\delta\rho (-{\bf q})\rangle\equiv S(q)=
\frac{m^2 n_0}{1-n_0 C(q)} \\
&&\langle\delta\rho ({\bf q})\delta n (-{\bf q})\rangle =-\frac{B}{h^{\prime
\prime}(\bar{n}_d)}S(q)b(q) \\
&&\langle\delta n ({\bf q})\delta n (-{\bf q})\rangle =\frac{k_B T}{h^{
\prime\prime}(\bar{n}_d)} ,
\end{eqnarray}
\label{stat:corr}
\end{mathletters}where $n_0\equiv\rho_0 /m$ is the
average number density. The first equation
reproduces the correct relation between the static structure factor $s(q)$
and the
direct correlation function $C(q)$. In calculating Eq.~(\ref{stat:corr}),
we neglected the
terms of order $B^2/\mu$ compared to those of order
$B/\mu$,
which is consistent with the self-consistent
condition obtained in Ref.~\cite{my2} for the slowing down of
the dynamics: The slowing down corresponds to
a weak coupling
($B\rightarrow 0$) and a shallow metastable well
for defects ($\mu\rightarrow 0$) with the condition $B\sim \mu^2$. As
indicated in Sec.~\ref{sec:1},
the coupling energy term is constructed such that
the static defect-density correlation function behaves like
$s(q) -1$. This is achieved in Eq.~(\ref{stat:corr}) by taking
\begin{equation}
b(q)=-n_0 C(q).
\label{bc}
\end{equation}
As one can see from Eqs.~(\ref{ry:eq}), (\ref{fc})
and (\ref{bc}),
the coupling term  $F_c$ has the same form as the quadratic
interaction term between the density fluctuations.

The construction of the generalized Langevin equations for
a general set of slow variables is described in
Ref.~\cite{mm}. The main ingredients in the
construction are the effective
Hamiltonian and the Poisson brackets between the slow
variables which are related to the streaming velocities governing
the reversible dynamics. In the present case, the set of slow variables
is given by $\{\rho , {\bf g}, n_d\}$ with the effective
Hamiltonian, Eq.~(\ref{f}). The Poisson brackets involving
the defect density $n_d$ are taken to have the same structure
as those involving $\rho$, since $n_d$ is a scalar quantity.
The calculation of the Langevin equations
in this case from the effective Hamiltonian
Eq.~(\ref{f}) is a straightforward generalization
of the wavenumber-independent case described in Ref.~\cite{my2,ym}.
The final expressions for the Langevin equations, however, are quite
complicated due to the presence of
the spatially-varying coefficients in the
effective Hamiltonian, which will
not be presented here. One can put those equations into
the field theoretic formalism of Martin-Siggia-Rose \cite{msr}, which
allows one to systematically calculate the nonlinear corrrections to
the Gaussian response and correlation functions. The detailed
calulation of the nonlinear corrections relevant to the glass transition
problem also follows closely the wavenumber-independent case
and the reader is referred to Ref.~\cite{my2,ym}.

In the framework of MCT, nonlinear couplings between density
fluctuations in the
dynamic viscosity are responsible for the dynamical slowing down.
This mechanism is most easily seen through the
normalized density auto-correlation function:
\begin{equation}
\phi ({\bf q},t)=\langle\delta\rho ({\bf q}, t)\delta\rho (-{\bf q},0)\rangle /
S(q).
\end{equation}
The Laplace transform of $\phi ({\bf q},t)$, defined by
$\phi ({\bf q},z)=-i\int^\infty_0 dt\exp (izt) \phi ({\bf q},t)$,
is represented as
\begin{equation}
\phi ({\bf q},z)=\frac{\rho z+iq^2\Gamma ({\bf q},z)}
{\rho [z^2 -\Omega^2_0 (q)]+izq^2 \Gamma ({\bf q},z)},
\label{phqz}
\end{equation}
where $\Omega^2_0 (q)$ is the microscopic phonon frequency
and $\Gamma ({\bf q},z)$ is the renormalized dynamic viscosity.

As discussed in
Sec.~\ref{sec:1},
MCT predicts that $\phi ({\bf q},t)$, as a function of time,
exhibits a very elaborate relaxation sequence.
As a first step, we restrict the
present analysis to the
intermediate time regime where the MCT analysis is significantly
simplified in the following aspects.
As shown in Ref.~\cite{goe1}, all the structural details of
the system in the intermediate time regime can be described by a
single number $\lambda$ which is related to the power-law
and the von-Schweidler exponents
$a$ and $b$ in the usual way \cite{goe2}:
\begin{equation}
\frac{\Gamma^2 (1-a)}{\Gamma (1-2a)}=\lambda =\frac{\Gamma^2 (1+b)}
{\Gamma (1+2b)},
\label{abl}
\end{equation}
where $\Gamma$ denotes the gamma function.
Consequently, $a$ and $b$ are wavenumber independent,
while all the other parameters describing the time relaxation
sequence
depend on the wavenumber. Also, in the intermediate time regime,
the defect auto-correlation function can be regarded as a constant
in time due to its extremely long time scale \cite{ym}:
$\langle\delta n ({\bf q},t)\delta n (-{\bf q},0)\rangle=k_B T/h^{\prime
\prime}(\bar{n}_d )$. Thus,
the analysis of the system in the $\alpha$-relaxation regime
must include the time dependence of the defect auto-correlation function
as was analyzed in wavenumber independent case in Ref.~\cite{ym}.
In the intermediate time regime, this kind of complication does not
occur and the renormalized viscosity can be expressed in terms of the
density auto-correlation function alone.

For the calculation of the nonlinear contribution to
$\Gamma ({\bf q},z)$, we follow the field-theoretic calculation in
Ref.~\cite{ym}. We find, in the time regime before
the primary $\alpha -$relaxation, the dynamic viscosity is represented by
\begin{eqnarray}
\Gamma ({\bf q}, z)=\Gamma_0 &+&\int^\infty_0 dt \: e^{izt}
\int\frac{d^3{\bf k}}{(2\pi )^3}  \label{rep:visco} \\
&&~~~~~~~~~~~~~~\Big[ V^{(1)}({\bf q},{\bf k})
\phi ({\bf k},t)+V^{(2)}
({\bf q},{\bf k})\phi ({\bf k},t) \phi ({\bf q}-{\bf k},t)\Big] ,
\nonumber
\end{eqnarray}
where $V^{(1)}$ and $V^{(2)}$ are appropriate vertices to be evaluated and
$\Gamma_0$ is the bare viscosity governing the microscopic dynamics.
The appearance of
the term linear in $\phi ({\bf k},t)$ in Eq.~(\ref{rep:visco})
is a direct consequence of the constant defect auto-correlation
function in this time regime \cite{bong}.

The vertices $V^{(1)}$ and $V^{(2)}$ depend on the coupling constant $B$
and the parameters $\mu ,\nu$ describing the metastable potential $h(n_d)$
for defects. The basic picture obtained in Ref.~\cite{my2} is that
the slowing down corresponds to the weak coupling ($B\rightarrow 0$)
and the low activation barrier ($\mu\rightarrow 0$) limit with the condition
$B\sim\mu^2$. In this limit, the vertices are completely
described by the average number density $n_0$ and the following
two dimensionless parameters:
\begin{equation}
y\equiv\frac{1}{n_0}\frac{\nu}{k_B T},~~~~~~~\kappa\equiv
\nu^2\frac{x}{\mu^2},
\label{ykappa}
\end{equation}
where $x\equiv (B\rho_0\bar{n}_d)/(n_0 k_B T )$ is the dimensionless
parameter describing the scale of the coupling energy.
We recall that $\nu$
is the parameter proportional to
the third derivative of $h(n_d)$ at $n_d =\bar{n_d}$.
Thus, when the second derivative $\mu\rightarrow 0$,
the parameter $ y $ just represents the overall scale of the double well
potential. The parameter $\kappa$ is the appropriate ratio between the weak
coupling ($x$) and the low barrier ($\mu^2$) for defects.
We find
that for the effective Hamiltonian, Eq.~(\ref{f}), the
vertices are given by
\begin{mathletters}
\begin{eqnarray}
&&V^{(1)}({\bf q},{\bf k})=\frac{s({\bf q})}{n_0}\Big[ 2y\frac{1}{q}
U({\bf q},{\bf k})s({\bf k})s({\bf q}-{\bf k})+\kappa n_0 C({\bf k})s({\bf k})
\Big]+\cal{O}(\mu ) \\
&&V^{(2)}({\bf q},{\bf k})=\frac{s({\bf q})}{n_0}\Big[ \frac{1}{2q^2}U^2(
{\bf q},{\bf k})-y\frac{1}{q}U({\bf q},{\bf k})\Big] s({\bf k})s({\bf q}-{\bf
k})+\cal{O}(\mu ), \label{v2}
\end{eqnarray}
\label{v12}
\end{mathletters}where
\begin{equation}
U({\bf q},{\bf k})\equiv \frac{1}{q}\Big[ ({\bf q}\cdot{\bf k})n_0 C({\bf k})
+({\bf q}\cdot ({\bf q}-{\bf k}))n_0 C({\bf q}-{\bf k}) \Big] .
\end{equation}
We note that in the previous MCT studies of hard-sphere fluids
\cite{conv:mct,mct:hs,mct:hs2,mct:hs3}
only the quadratic contribution ($V^{(2)}$) with the first term
in Eq.~(\ref{v2}) was considered, which corresponds to the case
where $y=\kappa =0$ in this model.

\section{Analysis of the Model}
\label{sec3}

Eqs.~(\ref{phqz}) and
(\ref{rep:visco}) with the vertices Eq.~(\ref{v12}) completely
specify our model which incorporates the wavenumber dependence of
the system.
In this section, we apply the model to the case of
a hard-sphere fluid using a known approximate but realistic
hard-sphere structure factor.

\subsection{Basic Equations from General MCT Analysis}

We first discuss the relevant results from the
detailed MCT treatment \cite{goe1} of Eqs.~(\ref{phqz}) and
(\ref{rep:visco}) for a general set of vertices
$V^{(i)},i=1,2,...M$.
It was found that the microscopic structural details of the
system are summarized into a single parameter $\lambda$.
This parameter is related to the
the power-law and von-Schweidler exponents
$a>0$ and $b>0$ via Eq.(\ref{abl}).
The MCT analysis
of the coupled set of equations, (\ref{phqz}) and
(\ref{rep:visco}) indicates an ergodic-nonergodic type transition
as parameters describing the vertices approach critical
values. The nonergodic phase is characterized by
$\phi ({\bf q}, t\rightarrow\infty )=f({\bf q})$ with
$f({\bf q})>0$. In the present case where $M=2$, the nonergodicity parameter
$f({\bf q})$ is given as a nonvanishing solution of
\begin{equation}
\frac{f({\bf q})}{1-f({\bf q})}=\int\frac{d^3{\bf k}}{(2\pi )^3}
\Big[ V^{(1)}({\bf q},{\bf k} )f({\bf k})+V^{(2)}({\bf q},{\bf k} )f({\bf k})
f({\bf q}-{\bf k} )\Big] ,
\label{crit}
\end{equation}
while in the ergodic phase, only the trivial solution $f(q)=0$ to
Eq.~(\ref{crit}) can be found.
The exponent parameter $\lambda$ is
determined for the critical values of the parameters by
\begin{equation}
\lambda =\int\frac{d^3 {\bf q}}{(2\pi )^3}\int\frac{d^3 {\bf k}}{(2\pi )^3}
\hat{e}_c ({\bf q})V^{(2)}({\bf q},{\bf k})
(1-f_c ({\bf k}))^2 (1-f_c ({\bf q}-{\bf k}))^
2 e_c ({\bf k})e_c ({\bf q}-{\bf k}),
\label{lam}
\end{equation}
where $e_c ({\bf k})$ and $\hat{e}_c ({\bf k})$
are left and right eigenvectors of
\begin{equation}
W({\bf q},{\bf  k})=(1-f_c ({\bf k}))^2 \Big[ V^{(1)}({\bf q},{\bf k})+
2V^{(2)}({\bf q},{\bf k})f_c ({\bf q}-{\bf k})\Big]
\end{equation}
\begin{equation}
\int\frac{d^3{\bf k}}{(2\pi )^3}W({\bf q},{\bf  k})e_c ({\bf k})=e_c ({\bf q}),
{}~~~\int\frac{d^3 {\bf q}}{(2\pi )^3}\hat{e}_c ({\bf q})W({\bf q},{\bf  k})
=\hat{e}_c ({\bf k})
\end{equation}
with the normalization $1=\int d^3{\bf k}/(2\pi )^3
\hat{e}_c({\bf k})e_c ({\bf k})
=\int d^3{\bf k}/(2\pi )^3 \hat{e}_c({\bf k})e^2_c
({\bf k})(1-f_c ({\bf k}))$.

As mentioned in the previous section, MCT predicts a two-step relaxation
process in the intermediate time regime. After the microscopic time scale
$\tau_0$, the normalized density auto-correlation function $\phi
({\bf q}, t)$ shows the power-law decay which can be described as
\cite{mct:hs2,goe3}
\begin{equation}
\phi ({\bf q}, t)=f_c (q)+h(q)c_\epsilon \Big(\frac{t}{\tau_a }
\Big) ^{-a}.
\label{power}
\end{equation}
This is valid for the time regime $\tau_0 \ll t\ll \tau_a$, where
the characteristic time scale is given by
\begin{equation}
\tau_a\sim |\epsilon |^{-\frac{1}{2a}}.
\end{equation}
Here $\epsilon$ is the control parameter indicating how far the system
is from the transition and thus the ideal glass transition corresponds to
$\epsilon =0$. In Eq.~(\ref{power}), the coefficients
are given by $c_\epsilon
=\sqrt{|\epsilon |}$ and $h(q)=(1-f_c (q))^2 e_c (q)$.
For longer times $t\gg\tau_a$, the system undergoes the von-Schweidler
relaxation if $\epsilon <0$
\begin{equation}
\phi ({\bf q}, t)=f_c (q)-h(q)
A_2 \Big(\frac{t}{\tau_\alpha }\Big) ^b
\label{von}
\end{equation}
with a positive constant $A_2$. This mechanism is valid
for the time scale $\tau_a\ll t\ll \tau_\alpha $
before the $\alpha$-relaxation, where the characteristic
time scale for the $\alpha$-relaxation is given by
\begin{equation}
\tau_\alpha \sim |\epsilon |^{-\gamma},~~~~~~\gamma=\frac{1}{2a}+
\frac{1}{2b}.
\end{equation}
For $\epsilon >0$, $\phi ({\bf q}, t)$ decays to a constant,
$f(q)=f_c(q)+h(q)c_\epsilon /\sqrt{1-\lambda}$. We note
that the critical
nonergodicity parameter $f_c (q)$ and the coefficient $h(q)$
do not depend on the control parameter $\epsilon$.

\subsection{Analysis for a Hard-sphere Fluid}

We calculate the vertices given by Eq.~(\ref{v12}),
using the Perkus-Yevick solution \cite{oz}
for the hard-sphere structure factor, $s(q\sigma)$
with the Verlet-Weiss correction
\cite{vw},
where $\sigma$ is the sphere diameter.
For this system, it is more
convenient to use the packing fraction $\eta$ instead of the average
number density $n_0$, where $\eta=\pi\sigma^3 n_0 /6$.
The wavenumber integration is done with the mesh
size $N=300$ and the upper cutoff $\Lambda\sigma =50$.

As can be seen from Eq.~(\ref{v12}), our model for the dynamics of
hard-sphere fluids depends on the packing fraction $\eta$
and the metastability parameters $y$ and $\kappa$. We recall that
$y$ describes the overall scale of the double-well potential
for defects and $\kappa$ is the appropriate ratio between the weak
coupling and the low activation barrier as defined in
Eq.~(\ref{ykappa}). We first investigate whether the double-well
potential and the coupling between the defects and the density fluctuations
can be arranged to yield the slowing down of the dynamics.
In order to see this,
we solve Eq.~(\ref{crit}) iteratively for given $y$, $\kappa$ and
$\eta$. We find that for given packing fraction $\eta$, a nontrivial
solution $f(q)>0$ can be found and thus
the dynamical slowing down is obtained
for certain values of parameters
$y$ and $\kappa$. In fact, there exsists a critical surface in
the $y$-$\kappa$ space
that separates the nonergodic ($f(q) >0$) and the ergodic ($f(q)=0$)
phases. This critical surface can be found for various values of
the packing fraction
(See Fig.~1). This result is quite different from that
found in the previous MCT studies of the hard-sphere system
\cite{conv:mct,mct:hs,mct:hs2,mct:hs3},
which corresponds to the case where $y=\kappa=0$.
According to those studies,
the transition occurs at the critical packing fraction $\eta_c\simeq
0.525$ \cite{mct:hs}. This value is determined by the packing
fraction at which the first nonvanishing solution of Eq.~(\ref{crit})
appears when $\eta$ is increased.
The present analysis shows that the dynamical
slowing down is controlled by the
metastability parameters $y$ and $\kappa$ arranging themselves to be on the
critical surface so that the transition occurs over
a range of $\eta$ which contains $\eta_c$ of the conventional MCT.

For given packing fraction, the values of $\kappa $ on the critical
surface do not change very much as shown in
Fig.~1. As the packing fraction decreases, these critical values of
$\kappa$ increase. Since the hard-sphere system remains
a fluid for small enough packing fraction, the parameter $\kappa$
should be bounded from above.
The exact upper bound, however, is not determined in the
present analysis for reasons discussed below.
For high packing fractions $\eta\gtrsim 0.525$,
the critical surface does not
extend to the small $y$ region:
For $\eta\gtrsim 0.525$ and for $y$ smaller than some
value $y_{\rm min}$,
one always has a nonvanishing solution $f(q)$ to Eq.~(\ref{crit}).
The value of $y_{\rm min}$ increases with increasing $\eta$
(See Fig.~1).
We note that for small packing fractions ($\eta\lesssim 0.505$)
or for large $y$ ($y\gtrsim 1.5$), the contribution to $f(q)$
from the large wavenumbers become significant
as seen in Fig.~2 (a) and (b). This makes
the present iteration analysis of Eq.~(\ref{crit}) with the
numerical large wavenumber cutoff $\Lambda\sigma =50$ less
reliable for sufficiently small or large packing fractions.
However, for the range of packing fractions,
$0.51\lesssim\eta\lesssim 0.54$, the present analysis clearly
shows the slowing down of the dynamics.

For the parameters on the critical surface, one can determine
the critical nonergodicity parameter $f_c (q)$ and the coefficient
$h(q)$ as defined in Eqs.~(\ref{power}) and (\ref{von}).
In Figs.~2 and 3, $f_c (q)$ and $h(q)$ are shown for various
values of $\eta$ and $y$ along with the corresponding quantities
of the conventional MCT at $\eta=\eta_c$.
In the framework of MCT, approaching the critical values of the parameters
is represented
by the parameter $\epsilon\rightarrow 0$
(See Eqs.~(\ref{power}) and (\ref{von})).
The conventional MCT analysis of hard-sphere fluids
\cite{conv:mct,mct:hs,mct:hs2,mct:hs3} is based on the assumption
that the transition is actually controlled by the packing fraction
approaching its critical value:
$\epsilon\sim\eta-\eta_c$. Thus, in those analyses, the critical
nonergodicity parameter $f_c (q)$ and the coefficient
$h(q)$ are regarded as parameters
independent of $\eta$.
In the present model, however, the transition
occurs when the system becomes metastable with the low activation
barrier ($\mu\rightarrow 0$) and the weak coupling ($x\rightarrow 0$)
satisfying the condition $x/\mu^2\rightarrow\kappa $
(See Eq.~(\ref{ykappa})), while
the metastability parameters $y$ and $\kappa$ arranging themselves
to be on the critical surface in $y$-$\kappa$ space
shown in Fig.~1.
Thus, the control parameter in this case
is proportional to $\mu$.
Therefore, in contrast to the
conventional MCT, $f_c (q)$ and $h(q)$
depend on the parameter $y$ describing
the overall scale of the defect potential as well as
the packing fraction.
As one can see from Fig.~2 for fixed $y$, $f_c (q)$ shows a variation
with $\eta$ such that it has larger (smaller) values
for low (high) wavenumbers as the packing fraction
increases.

For given point on the critical surface, $(y, \kappa (y))$,
one can calculate the exponent
parameter $\lambda$ using Eq.~(\ref{lam}).
Thus, in this model $\lambda$ is determined as
a function of $\eta$ and $y$ as shown in Fig.~4.
Consequently, the exponents $a$ and $b$ characterizing
the intermediate time regime are not subject to a
particular value of packing fraction, instead they are
given as smooth functions of $\eta$ with one additional parameter
$y$ (See Fig.~5).
As a function of the packing fraction, $\lambda (\eta )$ generally
takes smaller values as one goes into the region
of either small or large packing fractions.
It should be noted in Fig.~4 that the values of $\lambda$ become
less reliable for large $y$ ($y\gtrsim 1.5$) and small
$\eta$ ($\eta\lesssim 0.505$) because of the reason discussed above
concerning the contribution from the large wavenumbers. It is
important to note that within the band of packing fractions
$0.51\lesssim\eta\lesssim 0.54$ where the dynamical slowing
down is obtained,
$\lambda (\eta )$ is not a constant as a function of $\eta$.

\section{Discussion and Conclusion}
\label{sec:conc}

In this paper, we have been able to show how the metastability parameters
describing the defects are incorporated in the
structural arrest of hard-sphere fluids. The basic picture
we obtain is that the transition is not a sharp one at
a special value of packing fraction, but rather the
transition occurs over a range of packing fractions as
the metastability parameters arranging themselves
to be on the critical surface as shown in Fig.~1.

In some sense, the present model just replaces the role
played by the critical packing fraction $\eta_c \simeq 0.525$ of the
conventional MCT of the hard-sphere system
\cite{conv:mct,mct:hs,mct:hs2,mct:hs3} in the slowing down
of the dynamics by a band of packing fractions,
$0.51\lesssim \eta \lesssim 0.54$, which include $\eta_c$.
However, our model has very different physical implications
from the one studied in the conventional MCT in the
following aspects: (1)
In the conventional MCT,
the exponent parameter $\lambda$
is defined only at the critical packing fraction
$\eta_c$, where its value is evaluated as $\lambda\simeq 0.758$
\cite{mct:hs2}. This value is somewhat larger than most of the values
of $\lambda $ found in the present model (See Fig. 4).
Correspondingly, the values of the power-law and von-Schweidler exponents
$a$ and $b$
in this model are larger than the conventional MCT values,
$a=0.301$ and $b=0.545$
More importantly,
since our model is not tied to
the notion of $\eta_c$, the exponent parameter $\lambda$ can be
defined over a range of $\eta$. Furthermore, as a function of $\eta$,
$\lambda$ is not a constant, instead, it shows a variation as
shown in Fig.~4. (2) Other physical parameters such as the
critical nonergodicity parameter $f_c (q)$ and the coefficient
$h(q)$ describing the time relaxation sequence also depend on
$\eta$ in our model, while $f_c (q)$ and $h(q)$ are defined only
at $\eta=\eta_c$ in the conventional MCT. These differences can
be summarized in terms of the control parameter $\epsilon$
of each model such that in the conventional
MCT, $\epsilon\sim
\eta -\eta_c$ with the well-defined $\eta_c$
and $\epsilon$ is proportional to the metastability parameter
$\mu$ in the present model.

Recently, there have been many experiments on colloidal
systems consisting of spherical particles
\cite{exp:hs1,exp:hs2,exp:hs3},
to which the MCT analysis
of hard-sphere fluids has been applied.
It is claimed in those experiments
that the critical packing fraction $\eta_c$ for the glass transition
could be identified. The obtained values of $\eta_c$
range from $0.555\sim 0.575$ for
suspensions of PMMA particles \cite{exp:hs1,exp:hs2} to
$0.636$ for microgels \cite{exp:hs3}. However, it is important
to note that the identification of $\eta_c$
in those experiments is carried out using somewhat
indirect methods: For the PMMA system, $\eta_c$ is
taken as the value of $\eta$
at which a homogeneously nucleated crystallization
does not occur. For the system studied in Ref.~\cite{exp:hs3},
$\eta_c$ is determined by fitting the experimental data
against MCT results Eqs.~(\ref{power}) and (\ref{von}) {\em assuming}
that the control parameter is $\epsilon\sim\eta -\eta_c$.
Thus, these experimental results should not be considered
as a conclusive evidence for the
existence of $\eta_c$, which is defined by
the packing fraction at which the density
auto-correlation function $\phi ({\bf q},t)$
shows a sharp ergodic-nonergodic transition, {\it i.e.}
$\phi ({\bf q}, t)$
decays to a finite quantity if $\eta >\eta_c$ and to zero otherwise
as $t\rightarrow\infty$.
As we shall show below, one could instead find
an evidence from those experimental data
that supports the basic picture described in this paper
where the glass transition
in the hard-sphere system is in fact not a sharp one
at the well-defined transition density, but it
results from the system becoming metastable.

As can be seen from the claimed values of $\eta_c$ in those
experiments, the values of packing fraction where the
dynamical slowing down is observed are generally higher than
the band of $\eta$, $0.51\lesssim \eta \lesssim 0.54$ obtained here
and also than $\eta_c\simeq 0.525$ of conventional MCT. This
discrepancy has usually been neglected in the conventional MCT studies of
the hard-sphere system \cite{mct:hs,mct:hs2,mct:hs3}.
In fact, in those analyses, only the difference $
\eta -\eta_c$ is used to compare with
the experimental data, where
$\eta_c$ of the conventional MCT is identified with
the claimed values in the experiments. This discrepancy
in $\eta$ values between MCT and experiments is usually
attributed \cite{mct:hs} to the fact that MCT, without
including the cutoff effect
\cite{dm}, tends to overestimate the tendency to freeze.
We believe that this explanation also holds in the present case where
$\eta_c$ is replaced by the band of packing fractions. We note that
the systems considered in those experiments do not have the
ideal hard-sphere type potential as their inter-particle interactions,
which might also
contribute to this discrepancy.

In order to test experimentally the present picture of the
dynamical slowing down
in the hard-sphere system, one must study the dependence of
the exponent parameter $\lambda$ and the critical
nonergodicity parameter $f_c (q)$ on the packing fraction $\eta$.
A variation of these quantities as a function of $\eta$ would
imply that the system
undergoes a smooth transition without the special
transition packing fraction. In the experiments mentioned above,
$\lambda$ and $f_c (q)$ are obtained by using
Eqs.~(\ref{power}) and (\ref{von}) to fit the intermediate
time relaxation data of $\phi ({\bf q}, t)$.
But, unfortunately, in most
cases \cite{exp:hs1,exp:hs2},
the analysis is carried out under the assumption that
$\epsilon\sim\eta -\eta_c$.
In particular, the value of $\lambda$
is not obtained from the experimental data, instead
the conventional MCT value $\lambda \simeq 0.758$ is directly applied to
the analysis of the data. Also the apparent
variations of $f_c (q)$ with $\eta$ observed in Ref.\cite{exp:hs1,exp:hs2}
are not systematically studied but neglected in the analysis.
Thus, in order to see the $\eta$-dependence
of $\lambda$ and $f_c (q)$, a more careful study of
the experimental data is needed that treats
$\lambda$, $f_c (q)$, $h(q)c_\epsilon$ and $\tau_a$ in
Eqs.~(\ref{power}) and (\ref{von}) as adjustable parameters without
assuming any particular form for the control parameter $\epsilon$.

In fact, this kind of analysis has been performed at one point of
the discussion in Ref.~\cite{exp:hs3}.
For fixed wavenumeber $q$,
one can indeed see variations of $\lambda$ and $f_c (q)$ values
as functions of $\eta$
(See Fig.~5 of Ref.~\cite{exp:hs3}) when
they are treated as free parameters.
We note that the amount of variation of the exponent values
is actually comparable to the present findings (See Figs.~4 and 5),
although there
are some differences between
the actual values of the exponents $a$ and $b$ and the range of
packing fractions over which the exponents are measured.
The authors of Ref.~\cite{exp:hs3} actually concluded that these
variations are not systematic ones and treated $\lambda$ and $f_c (q)$
as constants throughout their analysis. In order to see if
this kind of data really shows a dependence of the physical parameters,
$\lambda$ and $f_c (q)$
on $\eta$ as described in the
present analysis, which results from the metastable nature of the transition,
a more detailed study is needed covering a wide range of
values of the wavenumbers and packing fractions.

The present model predicts that $f_c (q)$ and $\lambda$
depend on $\eta$ with one additional parameter $y$ (See Figs.~2 and 4).
We note that $y$ is the dimensionless parameter which represents
the overall scale of the potential energy for the defects.
We interpret this metastability
parameter $y$ as something
that depends
on the microscopic details of the system and that may also have
the packing fraction dependence in it, which is not, however,
determined within
the present theory. We believe that the parameter $y$
could be specified in principle as a function of packing
fraction through
a comparison with more detailed experimental data showing
the dependence of $f_c (q)$ and $\lambda$ on $\eta$.

We emphasize that in this paper we are mainly concerned with
the qualitative picture of the effect of the defect structure
on the slowing down of the dynamics of the hard-sphere
systems, since the defect structure has been approximated in a
simplest way using Eq.~(\ref{bc}).
Nevertheless, our model provides an important picture
of smooth transition with the physical parameters such
as the exponents $a$ and $b$ governing the time relaxation sequence
and the critical nonergodicity parameter $f_c (q)$ depending smoothly
on the packing fraction. And this general picture might be robust
in more complex systems. An example of such system
where the exponents depend
smoothly on the temperature can be found in the experiment
in Ref.~\cite{dixon}.

We also note that we have focused on the intermediate time regime before the
primary $\alpha$-relaxation. In this time regime, the defect
auto-correlation function can be regarded as a constant in time
\cite{my2,ym} due to its long time scale. For the wavenumber
independent case, one could extend the FNH formalism by considering
the time dependence of the defect auto-correlation function as shown in
Ref.~\cite{ym}. One can in principle extend the present
wavenumber-dependent model to the primary
relaxation regime by considering the full time and wavenumber
dependences of the density and the defect auto-correlation functions
following Ref.~\cite{ym}.
The analysis of such model, however, will be a very difficult task.

\acknowledgements
I gratefully acknowledge Professor Gene F. Mazenko for
his suggestion that I work on this subject. I am especially
grateful for his encouragement and support.
This work was supported primarily by the
MRSEC Program of the National Science Foundation
under Award Number DMR-9400379.

\begin{figure}
\caption{
The critical surface in the $y-\kappa$ space for
various values of packing fraction $\eta$. The thick dashed
line is $y_{\rm min} (\eta )$. For $\eta\geq 0.53$,
the critical surface does not extend to the small $y$ ($y
\leq y_{\rm min}$) regions.
}
\end{figure}

\begin{figure}
\caption{
(a) The critical nonergodicity parameter $f_c (q)$ for $y=0.1$
and for various values of packing fraction (pf).
(b) The critical nonergodicity parameter $f_c (q)$ for packing
fraction $\eta =0.52$ and for various values of $y$.
The thick solid
line is $f_c (q)$ of the conventional MCT at $\eta=\eta_c \simeq 0.525$.
}
\end{figure}

\begin{figure}
\caption{
The coefficient $h(q)$ describing the power-law and
the von-Schweidler relaxations as expressed
in Eqs.~(\protect\ref{power}) and
(\protect\ref{von}) for $y=0.1$ and
for various values of packing fraction (pf).
The thick solid
line is $h (q)$ of the conventional MCT at $\eta=\eta_c \simeq 0.525$.
}
\end{figure}

\begin{figure}
\caption{
The exponent parameter $\lambda$ as a function of packing
fraction $\eta$ for various values of $y$. The star ($\star$)
represents the conventional MCT value $\lambda\simeq 0.758$
at $\eta =\eta_c\simeq 0.525$.
}
\end{figure}

\begin{figure}
\caption{
The power-law and von-Schweidler exponents $a$ and $b$
as functions of packing
fraction $\eta$ for various values of $y$.
}
\end{figure}


\begin{references}
\bibitem{fn:dnd} Because of the coupling term, the fluctuation
$\delta n_d$ depends also on $\delta\rho$.
See Ref.~\cite{my2}.
\end{references}
\end{document}